# Due-to-Heatwaves Faults in Urban Distribution System: An Identification Approach

Andrea Mazza, Haoke Wu

**Abstract**

Distribution system faults occurring during heatwaves (HWs) are not all caused by the HW itself: concurrent factors such as asset ageing, mechanical defects, soil contamination, and operational constraints contribute independently. Hence, indiscriminately attributing all HW-period faults to thermal stress overestimates system vulnerability and misleads asset-management decisions. This paper proposes a systematic framework to identify and quantify the subset of summer faults directly attributable to HW occurrences (denoted Due-to-HW faults), by distinguishing them from Due-to-Others faults. HW events are first characterised through the Excess Heat Factor index. A covariance-based attribution criterion is then developed to distinguish faults whose occurrence is statistically consistent with HW-driven thermal mechanisms from those attributable to independent causes. Complementing the attribution framework, a time-delay model is introduced to estimate the lag between the beginning of a HW and fault occurrence by maximising the normalised covariance between hourly temperature series and shifted fault-duration series. Applied to six years of operational data from a real MV distribution network, the simulation results show that Due-to-HW faults constitute a significant yet variable proportion of total summer faults, underscoring the non-negligible impact of HW occurrences on summer fault statistics. Beyond documenting the deterioration of fault rate and Mean Time Between Failures across all seasons, the analysis confirms that Time-Between-Failures distributions depart significantly from the exponential assumption, with direct implications for the applicability of Poisson-based reliability models to distribution systems subject to recurrent HW stress.

## 1 Introduction

Climate change increasingly affects daily life because of anomalous weather conditions impacting both society and infrastructures. Several documents show the impacts on Europe in terms of modifications on the water cycle (e.g., reduced precipitation, more intense rainfall events, and altered river flows) and the presence of extreme temperatures, which contribute to drought conditions and increase wildfire risk (European Environment Agency (EEA) 2017; European Parliament 2018). Among temperature-driven hazards, the growing persistence of high-temperature episodes has motivated the deeper and deeper study of heatwaves (HWs). The power sector is strongly hit by elevated ambient temperatures such the ones reached during HWs. For example, they may constrain hydropower availability (Solaun and Cerda 2019), reduce the performance of steam power-plant cooling systems (Sieber 2013), and, important aspect to be considered for the success of the ongoing energy transition, degrade photovoltaic conversion efficiency (Spertino et al. 2015).

With reference to power grids, sustained heat has been demonstrated to be one of the causes that led to an increase of the fault rate, particularly for underground Medium-Voltage (MV) distribution systems. The thermal performances of cables and cable joints are strongly dependent on the surrounding soil, which serves as the primary heat dissipation medium (Liu et al. 2022). During prolonged hot and dry conditions, soil moisture depletion reduces thermal conductivity, leading to an increase conductor and insulation temperatures with respect to the temperature registered in “normal” conditions by considering the same loading (Malmedal et al. 2016). It is worth noting that the exposition of distribution systems to sustained periods of elevated ambient and soil temperatures produces cumulative thermal stress, which differs fundamentally from short-duration temperature excursions (Atrigna, Buonanno, et al. 2021). Multi-day heat exposure reduces the thermal recovery capability of underground assets, accelerates insulation ageing in cables and transformers, and degrades cable joints and interfaces (Anders 2005). As a result, even when daily peak temperatures are similar, prolonged HWs can significantly erode thermal margins and increase the likelihood of failures that are causally linked to heat-induced mechanisms.

The literature addressing HW impacts can be broadly organized into three main research directions:

- *Component-level investigations:* these contributions focus on measurements, thermal/ageing modeling, and fragility characterization of the most vulnerable elements, such as MV cable joints (Pompili et al. 2020). Physics-inspired, vulnerability-based methodologies have been proposed to quantify short-term failure probabilities of cables and joints through dynamic thermal models that include ageing processes (Ciapessoni et al. 2025). Experimental work has studied partial-discharge activity in MV joints and how increasing temperature affects inception voltage (Pompili and Calcara 2024), and structured laboratory protocols have been introduced to assess dielectric performance under combined thermal and electrical stress and to identify early degradation signatures (Calcara et al. 2025). Finally, monitoring technologies enable the data acquisition required for health assessment, including wireless self-powered temperature monitoring for joints (Muñoz et al. 2023) and multi-parameter sensing installations on underground lines (Peretto et al. 2018).

- *Fault prediction:* these studies aim to forecast fault occurrence to support DSO asset management and proactive operations. For example, (Atrigna et al. 2021) and (Aghahadi et al. 2025) compare machine-learning models for predicting MV faults under HW conditions, ranging from data-centric analyses to operationally oriented frameworks targeting reduced response time during fault events.

- *System-level analyses:* These approaches aim to study resilience and reliability by statistically linking the fault occurrence to meteorological conditions and by analyzing the evolution of relevant indicators. E-distribuzione has reported resilience assessments since 2018, considering HWs and ice sleeves among the extreme events (Amicarelli et al. 2018). Urban-grid investigations have identified critical scenarios and HW-sensitive substations through topological analyses such as Dulmage–Mendelsohn decomposition (Bragatto et al. 2019). Data-driven studies have correlated faults with weather features, showing that maximum temperature and minimum humidity can separate fault behavior into distinct clusters (Zhang et al. 2019). The *EHF* has also been used to document the worsening of HW conditions (e.g., in Torino, Italy) and to support the attribution of faults to HW conditions, enabling to carry out a cost–benefit analys (A. Mazza et al. 2021). Further work has examined the growing relevance of non-reliability faults (NRFs), showing that Time-Between-Failures cumulative distributions progressively deviate from exponential behavior as the share of HW-related NRFs increases (Mazza et al. 2024). More recent contributions leverage GIS-based approaches for planning (Montalà-Palau et al. 2025), emphasizing the distinction between *hazard* (the event) and *vulnerability* (component susceptibility), and their evolution, in line with NREL methodologies (Anderson et al. 2019).

The research on the HW effects on distribution system was supported by dedicated actions of National Authorities, which issued specific Regulatory Acts, for monitoring and verification purposes: for example, in 2017 the Italian Authority for Electricity, namely ARERA, introduced procedures explicitly aiming to improve transmission and distribution resilience (ARERA 2017). Indeed, the most common cause investigated by the Distribution System Operators (DSOs) was the HW occurrence.

The recognition of the HW occurrence, becomes then fundamental to provide a proper response to the Authorities and trying to alleviate the impacts that HWs may have on reliable supply. In this paper, the *EHF*, introduced in (Nairn et al. 2009) and refined in (Tolika 2019), has been used, even because it have demonstrated its effectiveness with respect to other metrics, as for example shown in in (Mazza and Wu 2025). The faults occurring during a HW, however, are not necessarily all related to the HW occurence. In fact, distribution system reliability is influenced by multiple multiple factors, e.g., electrical loading, asset ageing, soil conditions, cooling limitations, environmental contamination, protection-system (wrong) settings, and human errors (Sturchio et al. 2015). These factors exist independently of extreme temperature events: then, attributing all faults observed during a HW period directly to HW impacts can lead to misinterpretation of failure mechanisms and biased assessment of system vulnerability. Without a clear conceptual distinction, fault statistics may overestimate the direct impact of HWs. This motivates the introduction of the concept of ***Due-to-HW Faults***, defined as faults whose occurrence is primarily driven by *HW*-induced thermal mechanisms, such as cumulative overheating, reduced soil thermal conductivity, and accelerated thermal ageing (Falabretti et al. 2020). By contrast, faults that occur during a HW and predominantly caused by independent factors—such as advanced asset ageing, mechanical defects, contamination, or operational constraints—are classified as ***Due-to-Others Faults***. Establishing this distinction enables more accurate failure attribution and an enhanced understanding of the fault statistics under extreme weather conditions, by allowing utilities to differentiate between vulnerabilities that can be mitigated through HW-specific operational strategies and those requiring long-term asset replacement or maintenance interventions.

Against this background, the specific contributions of this paper with respect to the existing literature—including the authors' own prior work (A. Mazza et al. 2021), (Mazza et al. 2024), (Mazza and Wu 2025)—are as follows:

1. Fault-level attribution framework. Earlier studies (Amicarelli et al. 2018)–(Mazza et al. 2024) examine HW–fault correlations at an aggregate, seasonal, or annual scale without distinguishing individual faults by cause. The present work introduces the concept of *Due-* *to-HW* faults and a covariance-based criterion capable of labelling each fault individually, enabling fault-level attribution rather than period-level statistics. This is the core methodological novelty of the paper. The criterion is grounded in the cumulative thermal-stress mechanisms of underground MV cable systems, rather than being a purely empirical rule, yet is acknowledged as a statistical proxy rather than a deterministic physical proof.

2. Covariance-based time-delay model. Whereas (Yuan et al. 2020) addresses maintenance scheduling with a time-delay concept, this paper applies and extends the idea specifically to HW–fault dynamics in a distribution-grid context, quantifying the thermal-response lag at hourly resolution after filtering out non-HW faults.

3. Multi-year system-level evidence. The framework is applied to six years (2019–2024) of operational data from an urban Italian MV network, documenting a deteriorating trend in fault rate and *MTBF*, and demonstrating that the KS-test rejection of exponential *TBF* distributions has direct implications for reliability modelling practice.

The reminder of the paper is as follows: Section 2 presents the proposed method for identifying *Due-to-HW* faults. After that, Section 3 shows a covariance-based model to estimate the time delay between the fault occurrence and the HWs. Section 4 includes the simulation results and analysis. Finally, Section 5 reports the concluding remarks of the study.

# 2 Identification Method of Due-to-HW Faults

To operationalise the concept of *Due-to-HW Faults*, a quantitative identification method is required. This section presents the mathematical model that links HW intensity to daily fault occurrences.

## 2.1 Theoretical Rationale of the Covariance Criterion

The attribution criterion proposed in this work rests on the following physical argument. In an urban underground MV network, HW-driven fault mechanisms operate through cumulative thermal stress: sustained elevated ambient and soil temperatures reduce soil thermal conductivity, raise conductor and insulation temperatures above the values expected under equivalent loading at normal soil conditions, accelerate dielectric ageing in cables and joints, and ultimately trigger failure when the accumulated thermal damage exceeds a component-specific threshold (Atrigna, Buonanno, et al. 2021), (Anders 2005), (Ciapessoni et al. 2025). Because these mechanisms are directly coupled to the intensity and persistence of the HW event, faults arising from them are expected to occur preferentially on days when the *EHF* is elevated, producing a positive statistical dependence between daily *EHF* values and daily fault counts.

By contrast, faults driven by HW-independent factors—advanced cable ageing, mechanical defects at joints, soil contamination, protection miscoordination, or human error — are distributed in time according to their own failure-rate processes, which are not systematically coupled to HW intensity. The presence of such faults in the fault-count time series therefore introduces noise into the *EHF*–fault relationship, diluting the covariance rather than strengthening it.

These two contrasting behaviours motivate the identification criterion formalised in the remainder of this section. Basically, a fault is classified as *Due-to-HW* if its presence in the fault-count time series strengthens the statistical dependence between *EHF* and daily fault counts (that is, if removing it would reduce the covariance), whereas a fault whose presence reduces that dependence is classified as *Due-to-Others*. This is a necessary condition for causal attribution, not a sufficient one: the criterion will correctly label thermally driven faults with high probability when HW intensity genuinely modulates the fault process, but may misclassify individual faults in high-noise years where system-wide stressors weaken the *EHF*–fault signal. The criterion should therefore be understood as a statistical proxy for thermal causality, consistent with the data-driven nature of the analysis, rather than a deterministic physical proof.

It is also important to note what the criterion does not claim: it does not assert that every *Due-to-HW* fault would be absent if no HW had occurred (ageing processes may have brought a given component close to failure regardless), but rather that the HW-induced thermal stress was the proximate trigger that advanced the failure to the observed date. This interpretation is consistent with the cumulative-ageing framework established in (Anders 2005) and (Ciapessoni et al. 2025), and aligns with the operational definition of HW-attributable fault adopted by Italian regulatory practice (ARERA 2017).

## 2.2 Identification model of Due-to-HW faults

HWs are commonly described as prolonged periods of abnormally high temperatures relative to local climatological conditions. For distribution system resilience analysis, such qualitative descriptions must be translated into quantitative indicators that consistently characterise HW presence, intensity, and persistence. Several temperature-based indices have been proposed in the literature. Among them, the *EHF*, originally introduced in (Nairn et al. 2009), has been widely adopted and validated in different climatic regions, including Greece (Tolika 2019) and Italy (Andrea Mazza et al. 2021). Alternative indicators, such as the Index of HWs (*IHW*) used by some distribution system operators (Pompili et al. 2021), have also been reported.

A comparative study in (Mazza and Wu 2025) shows that the *EHF* outperforms alternative indices in both HW-period identification and statistical correlation with fault occurrences. Accordingly, the *EHF* is adopted in this work as the primary metric for quantifying HW conditions.

The *EHF* is defined on a daily basis and integrates both long-term climatological characteristics and short-term temperature anomalies (Nairn, Fawcett, et al. 2009). The *EHF* for day $i$ is given by:

$$EHF_i = \max\left(0, EHI_{\mathrm{sig},i}\right) \cdot \max\left(1, EHI_{\mathrm{accl},i}\right)$$

where $EHI_{\mathrm{sig}}$ is the significance index and $EHI_{\mathrm{accl}}$ is the acclimatization index.

The significance index is defined as:

$$EHI_{\mathrm{sig},i} = \frac{T_i + T_{i-1} + T_{i-2}}{3} - T_{95}$$

where $T_i$ denotes the daily mean temperature on day $i$, and $T_{95}$ is the 95th percentile of daily mean temperatures computed over a multi-decade reference period.

The acclimatization index is defined as:

$$EHI_{\text{accl},i} = \frac{T_i + T_{i-1} + T_{i-2}}{3} - \frac{1}{30}\sum_{k=1}^{30} T_{i-k}$$

A day $i$ is classified as a HW day when $EHF_i > 0$ (Saxena et al. 2024).

Following the conceptual definition in Section 1, a necessary condition for a fault to be classified as *Due-to-HW* is that its occurrence contributes positively to the statistical dependence between HW intensity and fault counts. To quantify this dependence, covariance is employed as a measure of joint variability between daily *EHF* values and daily fault numbers.

Let the daily *EHF* values be arranged in a row vector $\boldsymbol{h} \in \mathbb{R}^{1\times n}$ and the daily numbers in a row vector $\boldsymbol{f} \in \mathbb{R}^{1\times n}$, where $n$ denotes the number of days under analysis. The covariance matrix is given by:

$$\boldsymbol{M}(\boldsymbol{h}, \boldsymbol{f}) = \begin{bmatrix} \text{Var}(\boldsymbol{h}) & \text{Cov}(\boldsymbol{h}, \boldsymbol{f}) \\ \text{Cov}(\boldsymbol{h}, \boldsymbol{f}) & \text{Var}(\boldsymbol{f}) \end{bmatrix}$$

The variance and covariance terms are computed as:

$$\text{Var}(\boldsymbol{h}) = \frac{1}{n-1}\sum_{i=1}^{n} (h(i) - \bar{\mu}_{\boldsymbol{h}})^2$$

$$\text{Var}(\boldsymbol{f}) = \frac{1}{n-1}\sum_{i=1}^{n} \left(f(i) - \bar{\mu}_{\boldsymbol{f}}\right)^2$$

$$\text{Cov}(\boldsymbol{h}, \boldsymbol{f}) = \frac{1}{n-1}\sum_{i=1}^{n} (h(i) - \bar{\mu}_{\boldsymbol{h}})\left(f(i) - \bar{\mu}_{\boldsymbol{f}}\right)$$

where $\bar{\mu}_{\boldsymbol{h}}$ and $\bar{\mu}_{\boldsymbol{f}}$ denote the mean values of $\boldsymbol{h}$ and $\boldsymbol{f}$, respectively.

Let $\xi_j^{(i)}$ denote the $j$-th fault occurring on day $i$, and collect all faults into the matrix

$$\boldsymbol{\Gamma} = [\, \boldsymbol{c}_1, \, \boldsymbol{c}_2, \, \dots, \, \boldsymbol{c}_n \,]$$

where each column vector $\boldsymbol{c}_i$ contains all faults observed on day $i$. The daily fault count vector is given by:

$$\boldsymbol{f} = \left[\, |\Gamma_{:,1}|, \, |\Gamma_{:,2}|, \, \dots, \, |\Gamma_{:,n}| \,\right]$$

Removing a single fault $\xi_j^{(i)}$ results in a modified fault vector $\boldsymbol{f}^{-\xi_j^{(i)}}$, where the $i$-th column of $\boldsymbol{f}$ is reduced of one element. A fault $\xi_j^{(i)}$ is classified as *Due-to-HW Fault* if its presence strengthens the covariance between **h** and **f**. Namely:

$$\mathrm{Cov}(\boldsymbol{h}, \boldsymbol{f}) - \mathrm{Cov}(\boldsymbol{h}, \boldsymbol{f}^{-\xi_j^{(i)}}) > 0$$

Conversely, a fault $\xi_j^{(i)}$ is classified as a *Due-to-Others Fault* if:

$$\mathrm{Cov}(\boldsymbol{h}, \boldsymbol{f}) - \mathrm{Cov}(\boldsymbol{h}, \boldsymbol{f}^{-\xi_j^{(i)}}) \leq 0$$

# 3 Covariance-Based Time-Delay Modeling

HWs affect not only the fault counts but also the temporal characteristics of individual faults, particularly their duration. Unlike fault counts, which respond primarily at a daily scale, fault duration often exhibits a response delayed to thermal stress accumulation with higher temporal resolution (Yuan et al. 2020). After filtering out all *Due-to-Others Faults* using the identification method presented in Section [Sec:IDENTIFICATION METHOD OF DUE-TO-HEATWAVE FAULTS], the remaining *Due-to-HW Faults* allow the time-lag structure between HW conditions and fault duration to be explicitly quantified.

The *EHF*, although effective for HW identification, is defined at a daily level and is therefore insufficient to characterise the thermal environment (Klimenta 2022) at the exact time of fault occurrence. Multiple *Due-to-HW faults* may occur within the same day, each experiencing different temperature conditions that cannot be captured by a single daily *EHF* value. To accurately align ambient temperature with fault duration, hourly temperature data corresponding to the fault occurrence time are employed. This choice reflects the different temporal granularities of the two response variables used across the two stages of the proposed framework: fault counts are inherently daily aggregates, making the daily *EHF* the natural paired index in Section [Sec:IDENTIFICATION METHOD OF DUE-TO-HEATWAVE FAULTS]; fault duration is instead a continuous variable attached to a specific clock time, so matching it against the hourly temperature recorded at that same instant is both statistically appropriate and physically meaningful.

Consistent with the notation introduced in Section 2, the durations of *Due-to-HW faults* are arranged in a row vector:

$$\boldsymbol{g} \in \mathbb{R}^{1 \times m}$$

and the corresponding hourly temperatures at the fault occurrence times are arranged in a row vector:

$$\boldsymbol{e} \in \mathbb{R}^{1 \times m}$$

where $m$ denotes the total number of *Due-to-HW faults*.

The joint variability between fault duration and hourly temperature is characterised by the covariance matrix:

$$\boldsymbol{\Psi}(\boldsymbol{g}, \boldsymbol{e}) = \begin{bmatrix} \mathrm{Var}(\boldsymbol{g}) & \mathrm{Cov}(\boldsymbol{g}, \boldsymbol{e}) \\ \mathrm{Cov}(\boldsymbol{g}, \boldsymbol{e}) & \mathrm{Var}(\boldsymbol{e}) \end{bmatrix}$$

The variance and covariance terms are computed as:

$$\mathrm{Var}(\boldsymbol{g}) = \frac{1}{m-1}\sum_{k=1}^{m}\left(g(t_k) - \bar{\mu}_{\boldsymbol{g}}\right)^2$$

$$\mathrm{Var}(\boldsymbol{e}) = \frac{1}{m-1}\sum_{k=1}^{m}\left(e(t_k) - \bar{\mu}_{\boldsymbol{e}}\right)^2$$

$$\mathrm{Cov}(\boldsymbol{g}, \boldsymbol{e}) = \frac{1}{m-1}\sum_{k=1}^{m}\left(g(t_k) - \bar{\mu}_{\boldsymbol{g}}\right)\left(e(t_k) - \bar{\mu}_{\boldsymbol{e}}\right)$$

where $g(t_k)$ and $e(t_k)$ denote the fault duration and hourly temperature at time $t_k$, respectively, and $\bar{\mu}_{\boldsymbol{g}}$ and $\bar{\mu}_{\boldsymbol{e}}$ are the corresponding mean values.

Let $\Delta t_k$ denote the time delay between the onset of a HW and the occurrence of the $k$-th *Due-to-HW fault*, as illustrated in Fig. 1. To capture the delayed thermal effect on fault duration, the fault-duration vector is systematically shifted along the time axis.

By applying a sequence of temporal shifts, a matrix:

$$\boldsymbol{G}^{\Delta} \in \mathbb{R}^{m\times(m-1)}$$

is constructed, where each column corresponds to a shifted fault-duration vector:

$$\boldsymbol{g}^{(t_k+\Delta t_k)} \in \mathbb{R}^{m\times 1}$$

as illustrated in Fig. 2. The covariance between each shifted dataset and the temperature vector $\boldsymbol{e}$ is then evaluated.

To enable consistent comparison across different shifts, the normalised covariance, equivalent to the Pearson correlation coefficient (Nahler 2009), is employed. The normalized covariance $C_{\mathrm{ov}}^{(\mathrm{norm})}$ ranges from $-1$ to 1 (Ledoit and Wolf 2022). The optimal time delay $\Delta t_k^{\mathrm{opt}}$ is obtained by solving:

$$\Delta t_k^{\mathrm{opt}} = \underset{\Delta t_k}{\mathrm{argmax}} C_{\mathrm{ov}}^{(\mathrm{norm})}(\boldsymbol{e},\, \boldsymbol{G}^{\Delta})$$

The resulting $\Delta t_k^{\mathrm{opt}}$ represents the most statistically consistent delay between the thermal environment and the observed duration of *Due-to-HW faults* for the corresponding analysis period. A strictly positive $\Delta t_k^{\mathrm{opt}}$ is consistent with the cumulative thermal-stress mechanism: heat must accumulate in the cable trench over a period of days before the insulation temperature reaches a critical threshold that increases fault severity. A value of $\Delta t_k^{\mathrm{opt}}$ approaching zero indicates a more immediate thermal coupling, which may arise when cable components are already operating close to their thermal limits and therefore respond rapidly to further temperature increases. Conversely, a large $\Delta t_k^{\mathrm{opt}}$ points to a prolonged soil desiccation process during sustained HW events, whereby an extended thermal-accumulation phase is required before failures are triggered. The specific values of $\Delta t_k^{\mathrm{opt}}$ obtained for each year of the study, together with their physical interpretation in the context of the observed HW characteristics, are discussed in Section 4.

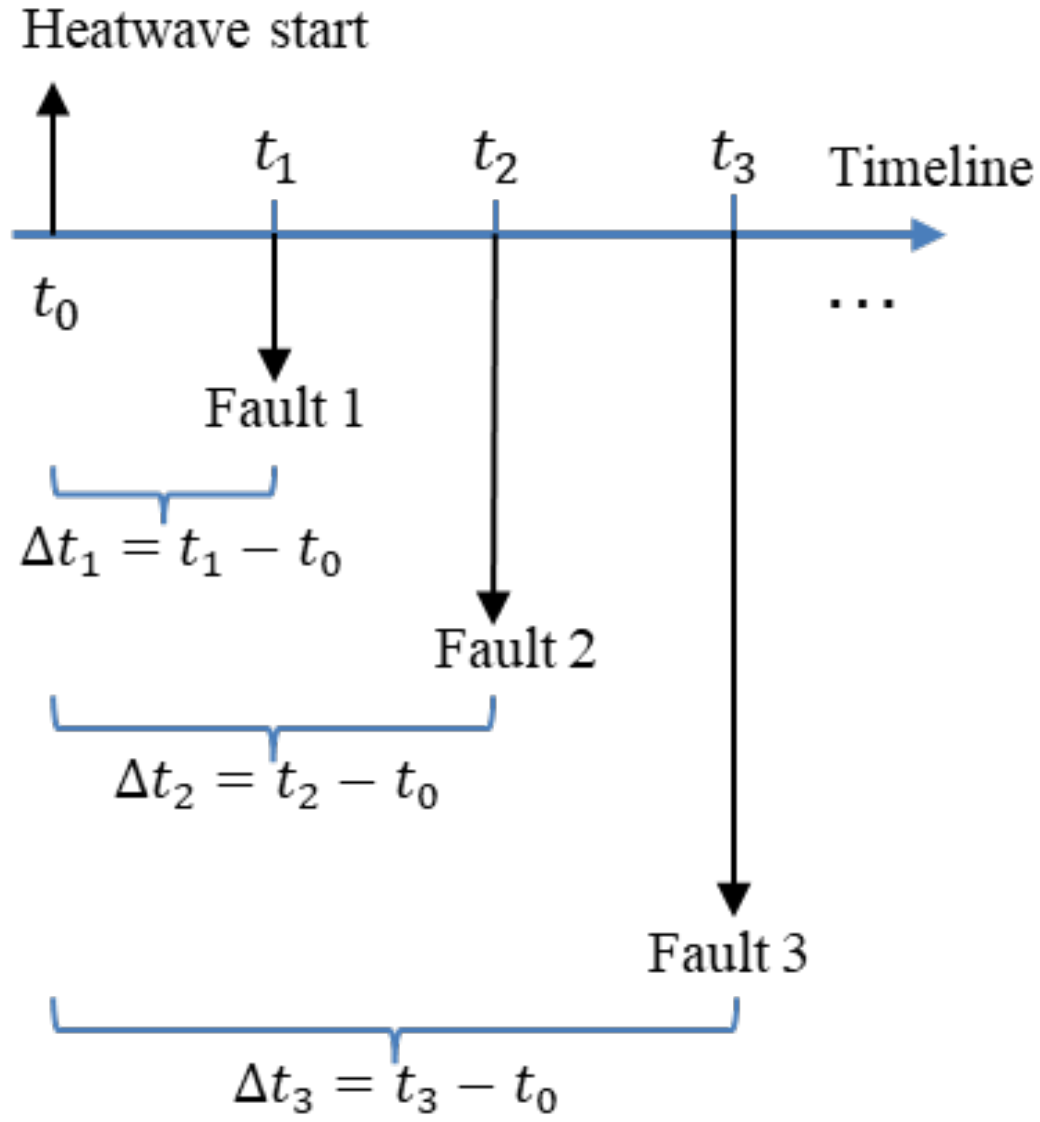

*Time delay between HW onset and individual fault occurrence.*

Hourly temperature $\mathbf{e}$
$e(t_0)$ $e(t_1)$ $e(t_2)$ $e(t_3)$
Timeline
$t_0$ $t_1$ $t_2$ $t_3$ ...
Shifted fault duration $\mathbf{g}^{(t_1+\Delta t_1)}$
$g(t_1)$ $g(t_2)$ $g(t_3)$ $g(t_4)$
Timeline
$t_1$ $t_2$ $t_3$ $t_4$ ...
Shifted fault duration $\mathbf{g}^{(t_m+\Delta t_m)}$
$g(t_m)$ $g(t_{m+1})$ $g(t_{m+2})$ $g(t_{m+3})$
Timeline
$t_m$ $t_{m+1}$ $t_{m+2}$ $t_{m+3}$ ...

*Illustration of fault-duration time shifting for covariance-based delay estimation.*

# 4 Simulation results and analysis

## 4.1 Algorithmic Implementation

The two stages formalised in Sections 2 and 3 are implemented as the single sequential algorithm summarised in Fig. 3. Block A performs the *Due-to-HW* fault identification loop: after computing the daily *EHF* series and the HW period, the algorithm iterates over every fault recorded within the HW period, recomputing $\mathrm{Cov}(\boldsymbol{h}, \boldsymbol{f}^{-\xi_j^{(i)}})$ after removing each fault in turn and applying the decision rule of Eqs. (8)–(9) to label it as *Due-to-HW* or *Due-to-Others*. Block B then takes the resulting *Due-to-HW* fault set, constructs the duration vector $\boldsymbol{g}$ and hourly-temperature vector $\boldsymbol{e}$, and searches the candidate-delay grid to determine the optimal lag $\Delta t_k^{\mathrm{opt}}$ that maximises the normalised covariance, as in Eq. (14). Both blocks are computationally inexpensive direct implementations, with Block A scaling with the number of faults tested and Block B scaling with the size of the delay search grid.

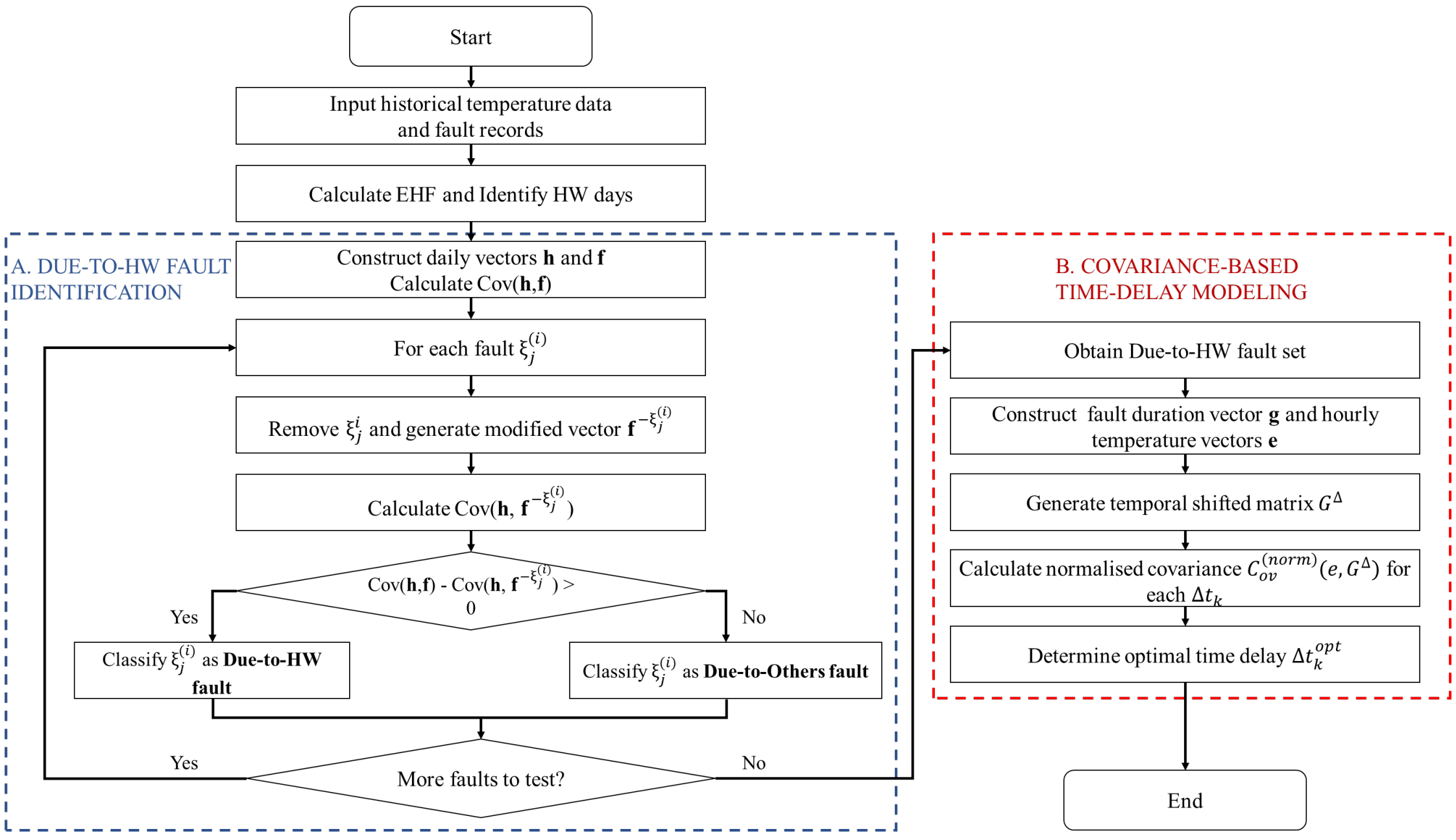

*Flowchart of the proposed methodology*

## 4.2 HW days identification

The daily minimum, average, and maximum temperatures of the City of Turin (North-Western Italy) are obtained from the website of the environmental agency ARPA Piemonte (ARPA Piemonte 2021). The $T_{95}$ value was calculated using daily temperature data from 1990 to 2018. This historical data enables the identification of *HW*s during the period from 2019 to 2024 using the *EHF*. The HW day is defined as the time when the *EHF* exceeds 0. As shown in Fig. 4, the year 2019 experienced the highest HW extreme, while years 2022 and 2024 had the longest HWs. In contrast, the year 2021 recorded the fewest HWs, with only two identified. As for the total number of HW days, 2022 recorded extreme HW events (42 days). In contrast, the year 2020 stands out for having very few HW days. Since the effects of HWs can extend beyond the identified HW days, this study defines the HW period as the interval from the initial HW day through the end of the summer month (September 30), while the non-HW period refers to the remaining days of the year, as shown in Table 1.

This definition is physically motivated: underground MV cables and their joints respond to ambient and soil temperature through a thermal time constant governed by the surrounding soil's volumetric heat capacity and thermal conductivity. During sustained hot-dry conditions, soil moisture depletion is a cumulative process that can persist for several weeks after the last

*EHF*-positive day, reducing the soil's heat dissipation capacity and keeping cable conductor temperatures elevated above the values expected at normal soil moisture levels for the same electrical loading (Malmedal et al. 2016), (Atrigna, Buonanno, et al. 2021). Defining the HW period as extending to September 30, i.e., the end of meteorological summer, captures this thermal aftereffect, which would otherwise be excluded if the HW period were terminated at the last $EHF > 0$ day. This definition is consistent with the approach adopted by ARERA (ARERA 2017) and with the prior work of the authors (A. Mazza et al. 2021), facilitating regulatory comparability.

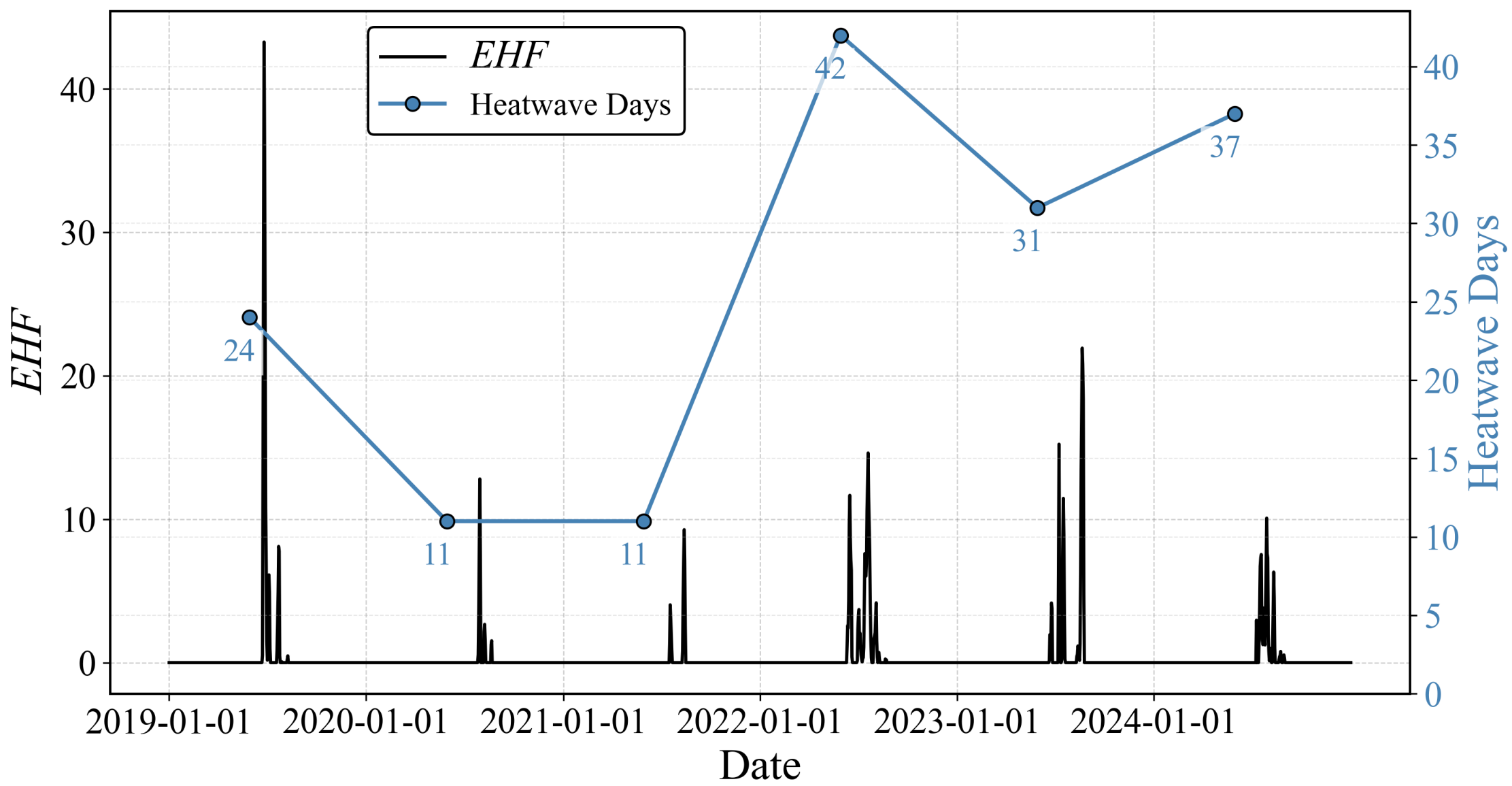

*HWs by EHF during 2019-2024.*

*Number of HW period and Non-HW period Days (2019–2024)*

| Year | HW Period (days) | Non-HW Period (days) |
|---|---|---|
| 2019 | 100 | 265 |
| 2020 | 66 | 300 |
| 2021 | 76 | 289 |
| 2022 | 112 | 253 |
| 2023 | 102 | 263 |
| 2024 | 84 | 282 |

## 4.3 Due-to-HW faults identification

The filtering of the faults from 2019 to 2024 was conducted using data referring to the distribution system of the city of Turin. By identifying *Due-to-HW faults*, *Due-to-Others faults* can be effectively filtered out. The results are presented in Fig. 5, which combines the annual counts of total faults and *Due-to-HW faults* with the covariance between the *EHF* and fault numbers, both before and after applying the proposed filtering model.

From 2019 to 2023, a sharp increase in total faults is observed, peaking in 2023 with 172 faults. *Due-to-HW faults* constituted a significant yet variable proportion of the total, ranging from approximately 40.5% (2021) to nearly 50.0% (2022), averaging 46.8% over the five years. The covariance results demonstrate a consistent and significant improvement after applying the proposed filtering method. Before filtering, the raw covariance values $Cov$(h,f) were significantly low, with an annual average of 0.288. The lowest value was recorded in 2023 (0.214), which coincided with the year of maximum fault count, indicating substantial noise in the weak correlation during high-stress periods.

After filtering, the *EHF*-fault covariance was substantially enhanced in every year, with a higher annual average of 0.810. The largest absolute improvement is observed in 2021, where the covariance jumps from 0.275 to 0.937—an increase of 0.662. This represents an improvement of over 240% relative to the baseline, highlighting the model's exceptional ability in isolating the *HW* signal during that period.

The year 2021 recorded the lowest number of total faults (42) and *Due-to-HW faults* (17), yet achieved the highest post-filtering covariance. The 2023 data shows the lowest post-filtering covariance (0.614) due to extreme system-wide fault noise, reaching an improvement of 0.4 from its pre-filter state. This suggests that during periods of lower overall system stress, the relationship between HWs and faults is inherently more discernible, and the filtering algorithm performs optimally. Also under high-noise conditions, the method still provides a measurable improvement.

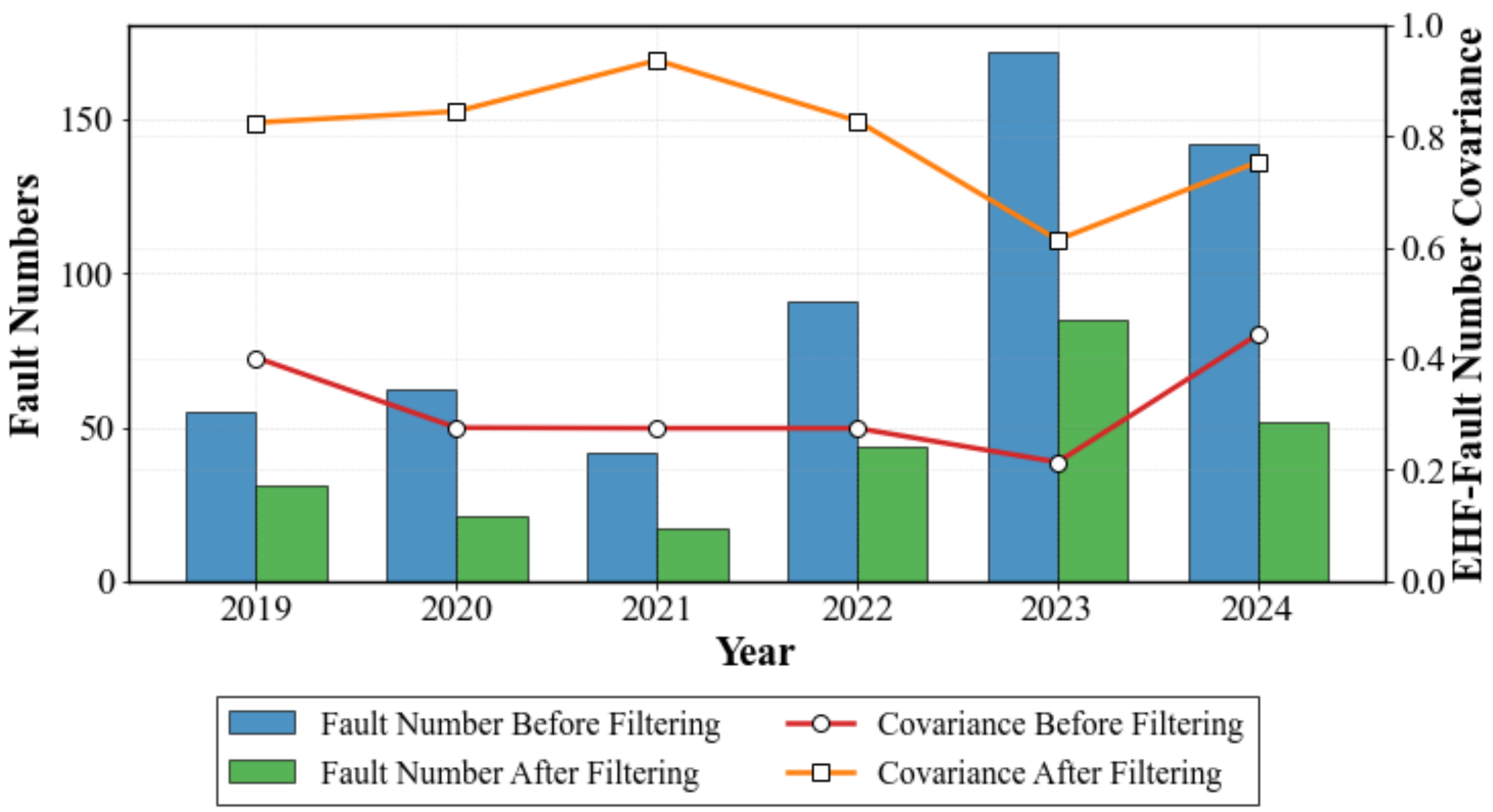

*Comparison before and after filtration.*

## 4.4 Time-delay calculation

Table [tab:covariance_comparison] summarises the alignment analysis between hourly temperature and fault duration for the period 2019–2024. For each year, the fault-duration time series is temporally shifted by an optimal lag $\Delta t^{\text{opt}}$ that maximises its normalised covariance

with the hourly temperature signal, thereby characterising the delayed thermal response of the system.

It should be noted that an increase in covariance after filtering is, to some extent, an expected outcome of the method itself, since the *Due-to-HW* classification is constructed precisely to strengthen the *EHF*–fault statistical dependence. The primary validation interest of this analysis therefore does not lie in the covariance improvement per se, but rather in whether the *optimal time-delay structure* recovered after filtering is physically consistent with the thermal-response mechanisms of underground MV cable systems. Specifically, if the filtering correctly isolates thermally driven faults, the resulting $\Delta t^{\mathrm{opt}}$ values should (a) be strictly non-negative, reflecting causal ordering from HW onset to fault occurrence; and (b) decrease in more recent years if the network's thermal margins have progressively eroded, causing faster fault responses to thermal exceedance.

Across all six years, the pre-filtering analysis consistently yields relatively low normalised covariance values $C_{\mathrm{ov}}^{(\mathrm{norm})}(\boldsymbol{e},\ \boldsymbol{G}^{\Delta})$ (ranging from 0.214 to 0.402), indicating that non-thermal disturbances largely obscure the temperature–fault duration relationship in the raw data. After filtering, the normalised covariance is substantially enhanced in every year, with improvements exceeding 100% in all cases. As discussed above, this improvement is partly methodologically expected; its relevance lies in confirming that the *Due-to-HW* subset retains a coherent temperature signal, whereas the removed *Due-to-Others* faults do not.

More specifically, in 2019 the optimal lag increases from 1.08 to 4.71 days, accompanied by a covariance enhancement of 105.22%, revealing a delayed response that is not evident in the raw data. Similar behaviour is observed in 2020 and 2021, where post-filtering covariances rise to 0.694 and 0.937, with a reduction of the optimal lag, suggesting a more immediate and stronger thermal coupling once non-HW faults are removed. In 2022, the optimal lag extends markedly to 11.25 days after filtering, indicating a pronounced long-term accumulation effect under sustained thermal stress. By contrast, in 2023 and 2024, the post-filtering optimal lags reduce to below one day, while covariance remains high (0.614 and 0.807), implying a faster system response under increasingly acute HW conditions.

For a more intuitive illustration, the year 2021 is selected as a representative example, and the corresponding alignment results before (Fig. 6) and after (Fig. 7) filtering are visualised to demonstrate the effect of the proposed methodology. In the unfiltered case the temperature and fault-duration series show weak and temporally incoherent alignment; after filtering, a better-defined lag structure emerges, with the shifted *Due-to-HW* fault-duration series aligning markedly better with the hourly temperature signal. Overall, the recovered time-delay patterns are physically meaningful and internally consistent across years, providing indirect but substantive support for the validity of the attribution framework beyond the self-referential covariance metric.

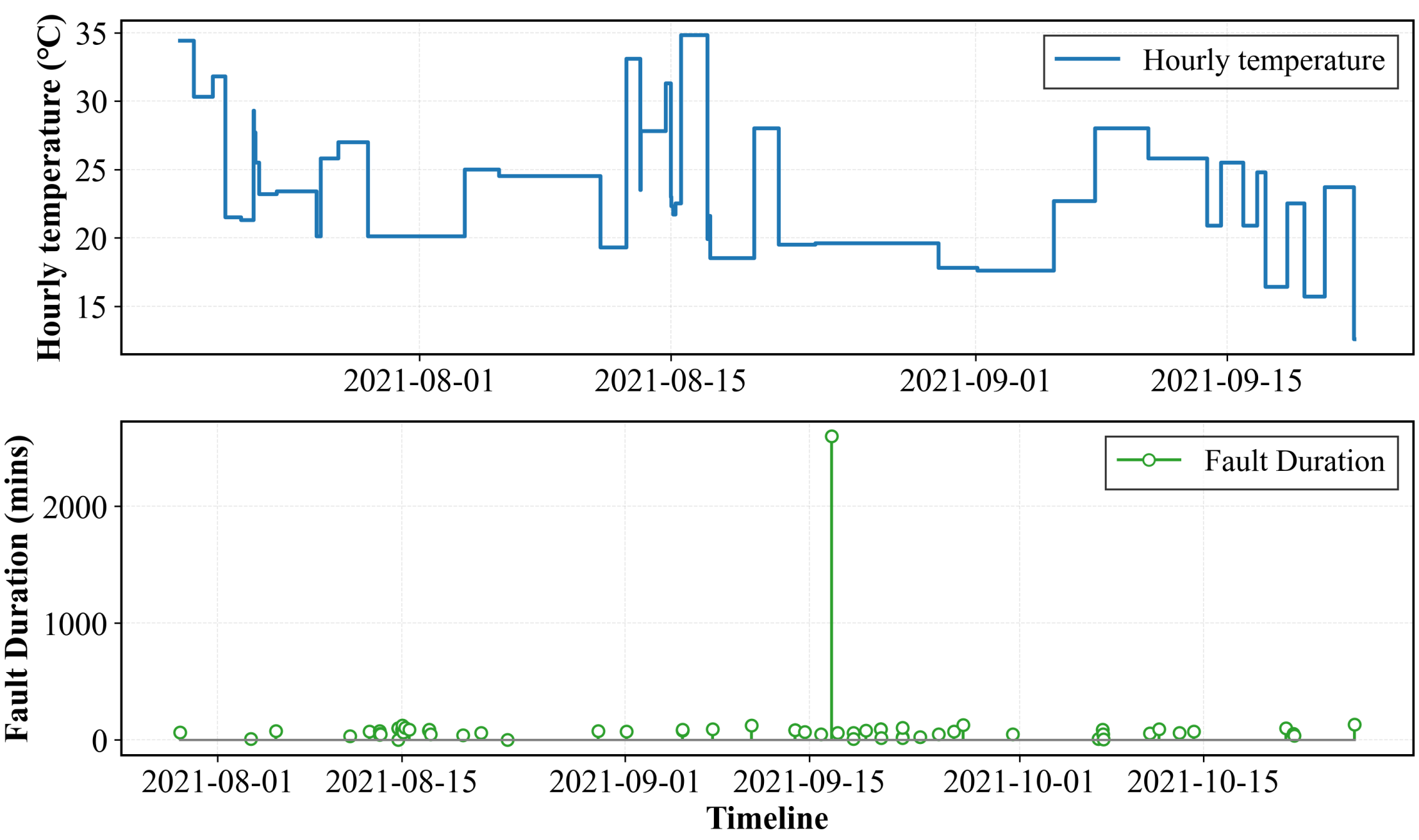

*Hourly temperature and All Fault duration values: the two timelines are shifted according to 2021 optimal Time Delay (10.49 days)*

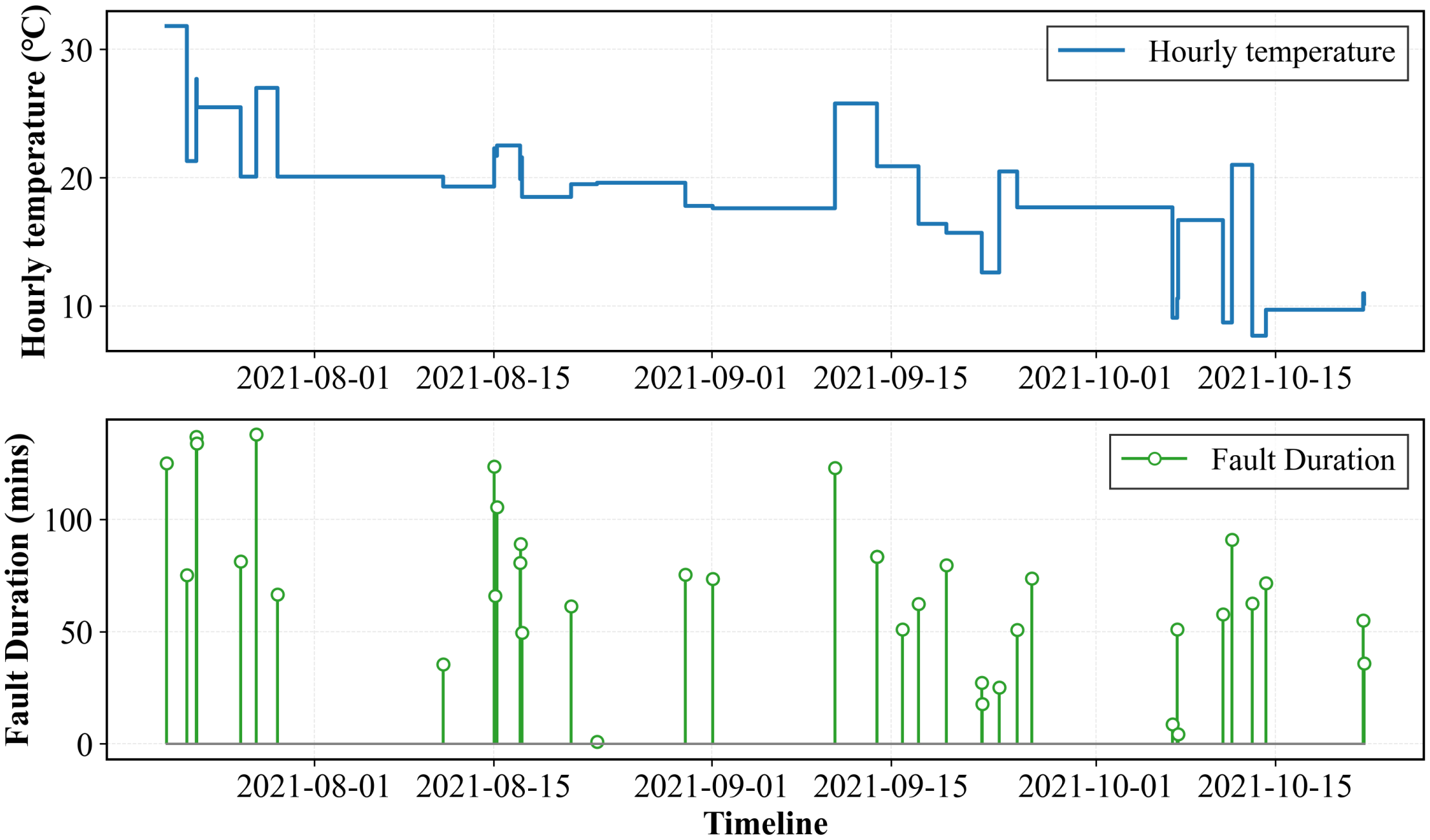

*Hourly temperature and Due-to-HW Fault duration values: the two timelines are shifted according to 2021 optimal Time Delay (3.48 days)*

## 4.5 Practical implication for DSOs

As shown in the following sections, the results obtained have practical implications on DSOs' planning and operation, since they directly impact the classification of the faults (and hence the fault rate calculation), as well as the simulation approaches to be used for reliability studies.

### 4.5.1 Fault rate and *MTBF* Analysis

Fault rate and *MTBF* are complementary reliability indices that jointly characterise the operational performance of underground cable systems (Hariharasubramanian 2025). The fault rate, expressed in [faults / (km·months)], quantifies the intensity of failure occurrence, while *MTBF*, expressed here in [days], reflects the average time interval between successive failures and thus represents overall system robustness.

| Period | Fault number | Line length (km) | Number of Months | Fault rate (fault/(km·months)) |
|---|---|---|---|---|
| 2019 winter | 41 | 1200 | 7 | 0.0049 |
| 2020 winter | 74 | 1225 | 7 | 0.0086 |
| 2021 winter | 120 | 1250 | 7 | 0.0137 |
| 2022 winter | 112 | 1275 | 7 | 0.0125 |
| 2023 winter | 117 | 1300 | 7 | 0.0129 |
| 2024 winter | 109 | 1325 | 7 | 0.0118 |
| | 49 | 1200 | 5 | 0.0082 |
| | 136 | 1225 | 5 | 0.0222 |
| | 126 | 1250 | 5 | 0.0202 |
| | 185 | 1275 | 5 | 0.0290 |
| | 176 | 1300 | 5 | 0.0271 |
| | 200 | 1325 | 5 | 0.0302 |
| 2019 summer | 84 | 1200 | 5 | 0.0140 |
| 2020 summer | 165 | 1225 | 5 | 0.0269 |
| 2021 summer | 154 | 1250 | 5 | 0.0246 |
| 2022 summer | 236 | 1275 | 5 | 0.0370 |
| 2023 summer | 271 | 1300 | 5 | 0.0417 |
| 2024 summer | 270 | 1325 | 5 | 0.0408 |

| Year | Winter | Summer excluding *Due_to_HW* faults | Summer |
|---|---|---|---|
| 2019 | 3.29 | 4.83 | 1.80 |
| 2020 | 2.91 | 1.71 | 1.07 |
| 2021 | 1.93 | 2.07 | 1.08 |
| 2022 | 1.88 | 1.19 | 0.68 |
| 2023 | 1.83 | 1.27 | 0.59 |
| 2024 | 1.84 | 1.07 | 0.58 |

From 2019 to 2024, the overall summer fault rate exhibits a sharp increasing trend, rising from 0.014 in 2019 to 0.0408 fault/(km·months) in 2024, accompanied by a corresponding reduction in *MTBF* from 1.80 to 0.58 days, as shown in Table [Fault rate statistics] and Table [MTBF statistic]. This deterioration is substantially more severe than that observed during winter, when the fault rate increases more moderately and the *MTBF* remains above 1.8 days in all years.

When *Due-to-HW* faults are excluded, the summer reliability metrics improve markedly. The new fault rate obtained by excluding *Due-to-HW* faults lies in the range 0.0082-0.0302 fault/(km·months), increasing from 2019 to 2024; in the meanwhile, the *MTBF* decreases from 4.83 to 1.07 days. Although a declining trend remains, the rate of degradation is significantly attenuated compared to the overall summer results. This practically means that, during the summer, the system would have been subjected to a deterioration of its reliability indexes even without HWs, but with lower magnitude.

Importantly, the gap between the overall summer *MTBF* and the *MTBF* excluding *Due-to-HW* faults widens over time, suggesting that the impact of HW events is not only significant but also intensifying. This observation aligns with the increasing frequency and severity of extreme temperature events in recent years.

A transverse comparison across seasonal categories further highlights the dominant influence of *Due-to-HW* faults. For all analyzed years, the overall summer fault rate is consistently 30–70% higher than that of summer excluding *Due-to-HW* faults, while the corresponding *MTBF* is reduced by more than 40% in most cases. For example, in 2022, the summer *MTBF* is 0.68, whereas excluding *Due-to-HW* faults increases it to 1.19, representing a 75% improvement.

Moreover, after excluding *Due-to-HW* faults, summer *MTBF* values in early years (e.g., 2019) exceed those observed in winter, indicating that under non-extreme thermal conditions, summer operation did not inherently exhibit lower reliability.

### 4.5.2 Analysis of the Empirical Cumulative Distribution Function (ECDF)

Using the calculated *MTBF*, *ECDF*s are generated for faults during the non-HW period, *Due-to-HW* faults during HW period and *Due-to-Others* faults during HW period. As shown in Fig.8, during non-HW periods, the *ECDF* for 2019 exhibits the best performance, followed by 2020, with minimal differences among the remaining years. For HW periods, considering only *Due-to-HW* faults, the *ECDF* performance from best to worst is 2019, 2021, 2020, 2022, 2024, and 2023 (Fig.9). In comparison, when only *Due-to-Others* faults are considered, 2019 again shows the best *ECDF* performance, followed by 2021 and 2022, while the differences among the remaining years are negligible, as shown in Fig.10, indicating a gradual worsening over time. A comparison between Fig.9 and Fig.10 further reveals that during the 2019 HW period, the maximum *TBF* for *Due-to-Others* faults exceeded 30 days, being substantially higher than that for *Due-to-HW* faults in the same period (indicating a better condition). However, this disparity progressively narrowed in subsequent years, eventually becoming negligible.

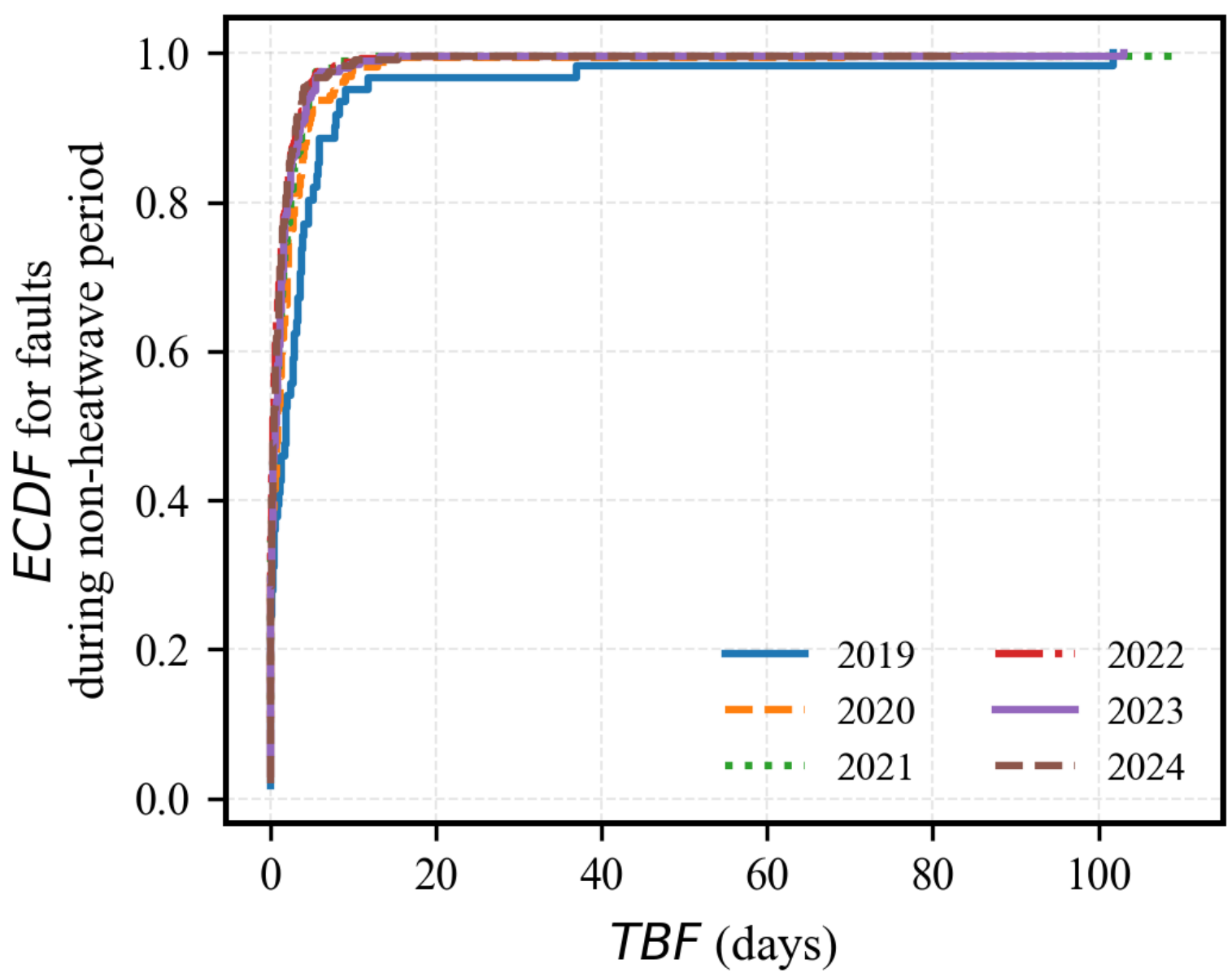

*ECDF for non-HW period*

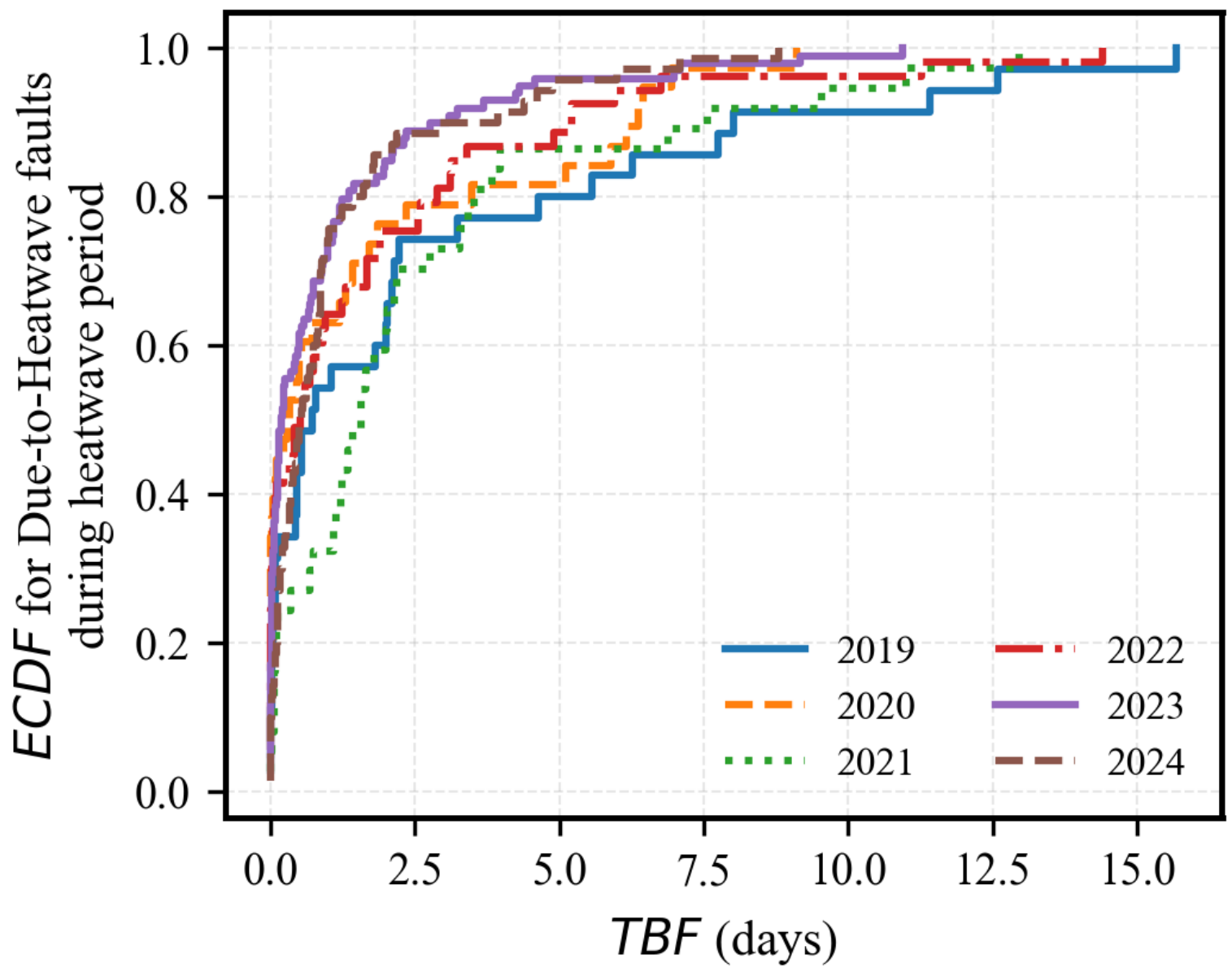

*ECDF for Due-to-HW faults*

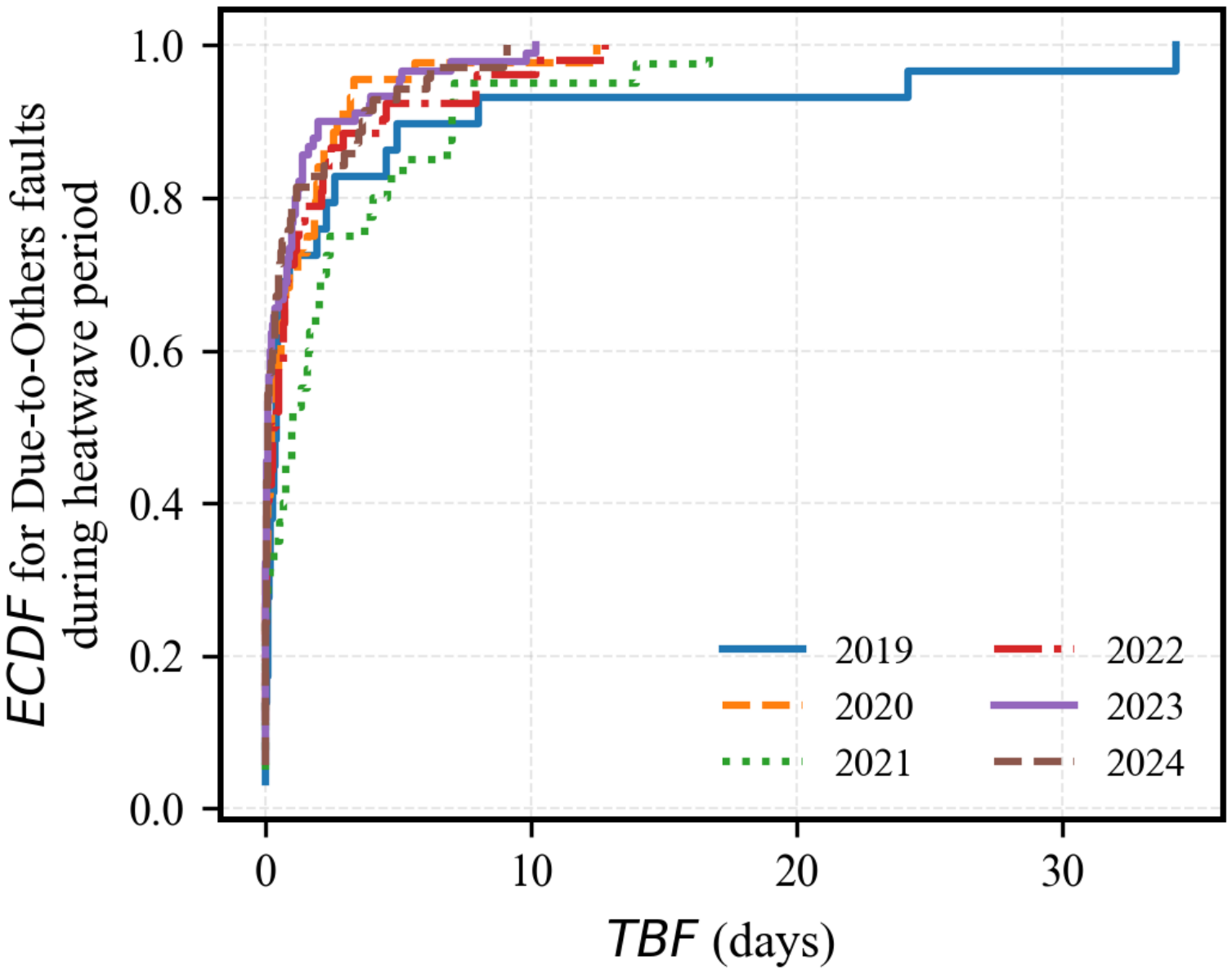

*ECDF for Due-to-others faults*

### 4.5.3 Characterization of the ECDFs

In the classical reliability theory, the *TBF*s follows a Poisson distribution $P(X=k)=\frac{\lambda^k e^{-\lambda}}{k!}, \quad k=0,1,2,\ldots$. When the system is reliable, the index value *k* is 0, causing the Poisson distribution to converge to an exponential distribution (Chowdhury and Koval 2009). To assess the conformity of the *ECDF* to an exponential distribution, we applied the Kolmogorov-Smirnov (KS) test (H. et al. 2000), in which the critical error threshold is determined based on the chosen significance level(0.01). The results are shown in Table [tab:ks test_results]: the hypothesis that *TBF* follows an exponential distribution is *rejected* for nearly all combinations (except for the winter of 2024). This has practical implications on how to model the fault occurrence in a distribution grid. Framework calculations exploiting the Poisson process (as, for example, in (Carpaneto and Chicco 2004)) cannot be straightly applied.

| **Period** | **D** | **Critical value** | **Result** |
|---|---|---|---|
| 2019 winter | 0.302 | 0.255 | Rejected |
| 2020 winter | 0.294 | 0.189 | Rejected |
| 2021 winter | 0.268 | 0.149 | Rejected |
| 2022 winter | 0.182 | 0.154 | Rejected |
| 2023 winter | 0.179 | 0.151 | Rejected |
| 2024 winter | 0.112 | 0.156 | Accepted |
| 2019 summer excluding *Due_to_HW* faults | 0.346 | 0.231 | Rejected |
| 2020 summer excluding *Due_to_HW* faults | 0.339 | 0.135 | Rejected |
| 2021 summer excluding *Due_to_HW* faults | 0.306 | 0.142 | Rejected |
| 2022 summer excluding *Due_to_HW* faults | 0.321 | 0.118 | Rejected |

| Period | D | Critical value | Result |
|---|---|---|---|
| 2023 summer excluding *Due*_to_*HW* faults | 0.325 | 0.121 | Rejected |
| 2024 summer excluding *Due*_to_*HW* faults | 0.315 | 0.114 | Rejected |
| 2019 summer | 0.328 | 0.178 | Rejected |
| 2020 summer | 0.327 | 0.127 | Rejected |
| 2021 summer | 0.294 | 0.131 | Rejected |
| 2022 summer | 0.338 | 0.106 | Rejected |
| 2023 summer | 0.336 | 0.099 | Rejected |
| 2024 summer | 0.284 | 0.099 | Rejected |

# 5 Conclusion

This paper proposes a framework for identifying *Due-to-HW* distribution system faults, with analysis of fault occurrences, timing delay, *MTBF*, and *ECDFs* under HWs.

A key contribution of this work lies in the systematic filtering of HW-related faults. The markedly lower covariance observed in raw operational data before filtering (average value of 0.288) demonstrates that unprocessed fault records contain substantial noise makes difficult recognizing climate–fault relationship. The proposed *Due-to-HW faults* identification model then becomes a prerequisite for revealing meaningful correlations between HWs and distribution system failures.

The validity of the identification framework is further supported by the physical consistency of the time-delay structure recovered after filtering: reducing to below one day in 2023–2024, consistent with the progressive erosion of the network's thermal margins evidenced by rising baseline fault rates—a physically coherent pattern that a misclassification-dominated result would not be expected to reproduce. Regarding the covariance improvement, in the lowest-stress year the post-filtering covariance reaches 0.937, the highest across the study period, while in the highest-stress year it remains at 0.614 despite a near-threefold improvement over the unfiltered baseline of 0.214, demonstrating that the framework recovers a meaningful thermal signal even under extreme noise dominance.

As expected, increasing fault rates are consistently associated with decreasing *MTBF* across years, confirming a deterioration in underground cable system reliability. From 2019 to 2024, the overall summer fault rate nearly tripled from 0.014 to 0.0408 faults/(km·months), with *MTBF* collapsing from 1.80 to 0.58 days. The *ECDF* analysis reveals a gradual deterioration in reliability over time as well, both in HW and non-HW periods. However, when *Due-to-HW* faults are excluded, summer *MTBF* values improve, surpassing those of winter in 2019 and 2021: hence, the summer operation does not certainly means a reliability deterioration in normal thermal conditions.

Another important point results from the KS test applied to the *ECDF*: the reject of the *TBF* exponential distribution hypothesis indicates that approaches making use of the Poisson process in the fault occurrence modeling should be carefully evaluated before being applied.

Despite its demonstrated effectiveness, several constraints should be acknowledged. KS test outcomes for low-count years are indicative rather than definitive, though a pooled test on the concatenated multi-year *TBF* series rejected the exponential hypothesis in all categories, supporting the year-by-year findings on a more robust basis. The covariance-based criterion is a statistical proxy for thermal causality, and its discriminating power is reduced in high-noise years. The case study covers a single operator in a single urban area, so framework parameters—particularly $T_{95}$ and the HW-period definition—require recalibration for other climatic contexts, though sensitivity analysis confirms the main conclusions are robust to the definitional choice adopted here. Future work will address this through cause-log annotation and application to additional distribution systems in different climatic regions.

## Acknowledgment

This study was developed within the project EXTRASTRONG (resilience evaluation by EXperimental and TheoRetical ApproacheS in electrical distributiON systems with underGround cables) – funded by European Union – Next Generation EU within the PRIN 2022 program (D.D.104 – 02/02/2022 – Ministero dell'Università e della Ricerca). This manuscript reflects only the authors' views and opinions, and the Ministry cannot be considered responsible for them.